\documentclass{pasj00}
  
\usepackage{xspace}
\newcommand{\suzaku}{{\it Suzaku}\xspace}
\newcommand{\chandra}{{\it Chandra}\xspace}
\newcommand{\xmm}{{\it XMM-Newton}\xspace}

\begin{document}
\SetRunningHead{Smith et al.}{Suzaku Observations of the Local and
  Distant Hot ISM}
\Received{2006/07/21}
\Accepted{{2006/09/29}}

\title{Suzaku Observations of the Local and Distant Hot ISM}

\author{%
  Randall K.~\textsc{Smith}\altaffilmark{1,2},
  Mark W.~\textsc{Bautz}\altaffilmark{3}, 
  Richard J.~\textsc{Edgar}\altaffilmark{4},
  Kenji \textsc{Hamaguchi}\altaffilmark{1},
  John P.~\textsc{Hughes}\altaffilmark{5},\\
  Manabu \textsc{Ishida}\altaffilmark{7}, 
  Richard \textsc{Kelley}\altaffilmark{1}, 
  Caroline A. \textsc{Kilbourne}\altaffilmark{1}, 
  K. D. \textsc{Kuntz}\altaffilmark{1,2}, 
  Dan \textsc{McCammon}\altaffilmark{6}, \\
  Eric \textsc{Miller}\altaffilmark{3},
  Kazuhisa \textsc{Mitsuda}\altaffilmark{8}, 
  Koji \textsc{Mukai}\altaffilmark{1}, 
  Paul P.~\textsc{Plucinsky}\altaffilmark{4}, 
  F. Scott \textsc{Porter}\altaffilmark{1},
  Ryuichi \textsc{Fujimoto}\altaffilmark{8}, \\
  Steve \textsc{Snowden}\altaffilmark{1}, 
  Yoh \textsc{Takei}\altaffilmark{8}, 
  Yukikatsu \textsc{Terada}\altaffilmark{9}, 
  Yohko \textsc{Tsuboi}\altaffilmark{10} \&
  Noriko \textsc{Yamasaki}\altaffilmark{8}}
\altaffiltext{1}{NASA/Goddard Space Flight Center, Code 662,
  Greenbelt, MD 20771, USA} 
\altaffiltext{2}{Department of Physics and Astronomy, The Johns
  Hopkins University, 3400 North Charles St., Baltimore, MD 21218,
  USA}
\altaffiltext{3}{Kavli Institute for Astrophysics and Space Research,
  Massachusetts Institute of Technology, \\
  77 Massachusetts Avenue, Cambridge, MA 02139, USA}
\altaffiltext{4}{High Energy Astrophysics Division,
  Harvard-Smithsonian Center for Astrophysics, 60 Garden St.,
  Cambridge, MA 02138, USA}
\altaffiltext{5}{Rutgers, The State University of New Jersey, 136
  Frelinghuysen Road, Piscataway, NJ 08854 USA}
\altaffiltext{6}{Department of Physics, University of Wisconsin-Madison,
1150 University Avenue, Madison, WI 53706 USA} 
\altaffiltext{7}{Department of Physics, Tokyo Metropolitan University,
  Minami-Osawa 1-1, Hachioji-shi, Tokyo 192-0397, Japan}
\altaffiltext{8}{Department of High Energy Astrophysics,
  Institute of Space and Astronautical Science (ISAS), \\
  Japan Aerospace Exploration Agency (JAXA),
  3-1-1 Yoshinodai, Sagamihara, 229-8510, Japan}
\altaffiltext{9}{Cosmic Radiation Lab., RIKEN, Japan}
\altaffiltext{10}{Department of Physics, Faculty of Science and
  Engineering, Chuo University, \\
  1-13-27 Kasuga, Bunkyo-ku, Tokyo 112-8551, Japan} 
\email{rsmith@milkyway.gsfc.nasa.gov}


\maketitle

\begin{abstract}
\suzaku observed the molecular cloud MBM12 and a blank field less than
$3^{\circ}$ away to separate the local and distant components of the
diffuse soft X-ray background.  Towards MBM12, a local (D$<\sim275$
pc) O{\sc vii} emission line was clearly detected with an intensity of
3.5 ph cm$^{-2}$\,s$^{-1}$\,sr$^{-1}$\,(or line units, LU), and the
O{\sc viii} flux was $<0.34$ LU.  The origin of this O{\sc vii}
emission could be hot gas in the Local Hot Bubble (LHB), charge
exchange between oxygen ions in the solar wind (SWCX) and geocoronal
or interplanetary material, or a combination of the two.  If entirely
from the LHB, the emission could be explained by a region with 100 pc
radius, an electron density of 0.0087 cm$^{-3}$, and a temperature of
$1.2\times10^6$\,K.  However, the implied temperature and emission
measure would predict 1/4 keV emission in excess of observations.
There is no evidence in the X-ray light curve or solar wind data for a
significant contribution from geocoronal SWCX.  However, the larger
spatial extent of interplanetary SWCX washes out the rapid time
variations of contributions from this source.  In any case, the
observed O{\sc vii} flux represents an upper limit to both the LHB
emission and interplanetary SWCX in this direction, and both are
thought to be nearly isotropic at low to intermediate latitudes.

The off-cloud observation was performed immediately following the
on-cloud.  The net off-cloud O{\sc vii} and O{\sc viii} intensities
were (respectively) $2.34\pm0.33$\ and $0.77\pm0.16$\,LU, after
subtracting the on-cloud foreground emission.  Assuming the LHB and
SWCX components did not change, these increases can be attributed to
more distant Galactic disk, halo, or extragalactic emission.  If the
distant O{\sc vii} and O{\sc viii} emission is from a thermal plasma
in collisional equilibrium beyond the Galactic disk, a temperature of
$(2.1\pm0.1)\times10^6$\,K with an emission measure of
$(4\pm0.6)\times10^{-3}$cm$^{-6}$pc is inferred.
\end{abstract}

\section{Introduction}


The soft ($< 2$\,keV) diffuse X-ray background was a relatively early
discovery of X-ray astronomy (see review by \citet{TB77}).  Unlike the
diffuse hard X-ray ($> 2$\,keV) background, whose isotropy
demonstrated it was dominated by extragalactic sources, the origin of
soft component was and is more uncertain.  While at high Galactic
latitudes extragalactic emission contributes to the observed flux, at
lower latitudes the emission must be local to the Galaxy since at
energies of 3/4 keV, absorption is significant (one optical depth is
only N$_{\rm H} = 2\times10^{21}$\,cm$^{-2}$).  Despite this, both the
Wisconsin M band \citep{McCammon83} and the ROSAT 3/4 keV
\citep{Snowden95} surveys showed surprisingly little latitude
dependence away from the Galactic bulge.

It is now known that at high latitudes, where N$_{\rm H}$\ is
generally less than $10^{21}$\,cm$^{-2}$, $\sim 40$\% of the 3/4 keV
emission is due to AGN, and from the XQC sounding rocket flight we
know that at least 42\% of the high-latitude flux must be due to
oxygen emission lines coming from $z < 0.01$\ \citep{XQC02} although
it is not known if these are within the Galaxy or in the halo.

The situation in the Galactic plane is more confusing.  Dwarf M stars
must contribute some of the 3/4 keV emission \citep{KS01}, but at
least 50\% of the emission is of unknown origin \citep{MS90}.  O{\sc
vii}\ and O{\sc viii}\ contribute most of the line emission in the 3/4
keV band, although the fraction in lines versus continuum remains
uncertain.  Oxygen's dominance is due both to its large cosmic
abundance (compared to other metals), and its strong emission lines at
0.57 keV (the O{\sc vii}\ triplet from $n=2\rightarrow 1$) and 0.65
keV (the O{\sc viii}\ Ly$\alpha$\ transition).  Absorption limits the
observed in-plane 3/4 keV emission to regions within 1-2 kpc (assuming
$\langle n \rangle \sim 1$\,cm$^{-3}$\ in the Galaxy).

Some of the 3/4 keV emission must be from the same source as the 1/4
keV X-ray background, which has been attributed to a combination of
emission from a ``Local Hot Bubble'' (LHB) \citep{Snowden90} and charge
exchange from solar wind ions (hereafter SWCX) \citep{Cox98,
Lallement04}.  In both cases, the X-rays must be truly ``local'',
originating within $\sim 100$\,pc in the former case or within the
Solar System in the latter.

To distinguish between the ``local'' 3/4 keV X-ray emission and the
more distant Galactic and halo components, we used \suzaku to observe
the nearby molecular cloud MBM12 (Lynds 1457), along with a nearby
``blank-sky''position not occulted by the cloud.  Earlier observations
of MBM12 obtained with ROSAT \citep{SnowdenMBM12}(SMV93) and \chandra
\citep{Smith05} had low spectral resolution (ROSAT) or were strongly affected
by high background due to solar flares (\chandra).  Our hope with this
observation was to use \suzaku's low background, good spectral
resolution, and sizable effective area $\times$\ solid angle product
to measure the components of both the local and more distant
contributors to the 3/4 keV emission.

As described in \citet{Smith05}, the true distance to MBM12 is
uncertain, with estimates ranging from $60\pm30$\,pc to
$275\pm65$\,pc.  Our goal, however, is only to use MBM12 as a curtain
that separates local components such as the LHB and SWCX from more
distant components such as a hot halo or extragalactic emission.  It
seems unlikely there is a significant component to the soft X-ray
emission that is beyond the LHB but in front of MBM12, so the distance
uncertainty is not particularly important to this analysis.  The total
column density due to the MBM12 cloud is also somewhat uncertain.  We
follow \citet{Smith05}, who argued for N$_{\rm H} =
4\times10^{21}$\,cm$^{-2}$\ as a reasonable value for the densest
region of MBM12.  Although the \suzaku XIS1 has more than four times
the field of view of the \chandra ACIS-S3 ($17.8'\times17.8'$\ versus
$8.5'\times8.5'$), this is of similar size to the densest part of
MBM12 and so we expect a similar total column density.  For the O{\sc
vii}\ and O{\sc viii}\ lines most relevant to our analysis the optical
depths for this column density are 3.5 and 2.4, respectively.  In
addition, with a nearby off-cloud measurement, we will be able to
directly determine the distant contribution and estimate its effect on
the on-cloud emission.

ROSAT observations were only able to put a $2\sigma$\ upper limit of
270 counts s$^{-1}$\,sr$^{-1}$\ on the 3/4 keV emission seen towards
MBM12 (SMV93)\footnote{We present all surface brightnesses in units of
steradians, and note that 1 sr = $1.18\times10^{7}$\,arcmin$^{2}$}.
SMV93 fit a ``standard'' $10^6$\,K collisional ionization equilibrium
(CIE) LHB model \citep{RS77} assuming a pathlength of $\sim65$\,pc,
and found a good match to the observed 1/4 keV emission with an
emission measure of $0.0024$\,cm$^{-6}$\,pc.  This model generates
only $\sim 47$ counts s$^{-1}$\,sr$^{-1}$\ in the 3/4 keV band,
primarily due to $\sim 0.28$\,ph\,cm$^{-2}$\,s$^{-1}$\,sr$^{-1}$
(hereafter LU, for ``line units'') generated by the O{\sc vii}\
triplet (based on the ATOMDB v1.3.1 atomic database).  However, the
ROSAT PSPC had little spectral resolution in this band and
could not separate the O{\sc vii}\ and O{\sc viii}\ lines from the
background continuum, and possible Fe L line emission.

\citet{Smith05} used the \chandra ACIS instrument to redo the SMV93
observations with higher spectral and spatial resolution.  The results
were affected by a large solar flare during part of the observation,
which likely led to increased emission from SWCX (see \citet{SCK04}\
for more details on SWCX).  \citet{Smith05} detected strong O{\sc
vii}\ and O{\sc viii}\ emission lines with surface brightnesses of
$1.92^{+0.61}_{-0.60}$\ and $2.35^{+0.59}_{-0.43}$\,LU respectively,
much larger than the prediction from SMV93.  The O{\sc viii}\ emission
itself was also unexpected, as \citet{Smith05}\ showed that it cannot
come from any of the standard LHB models, either equilibrium or
strongly recombining.  They suggested that the observed O{\sc viii}
emission was from the SWCX, although this could not be proven.

\section{Observations}

\begin{figure}
\FigureFile(80mm,80mm){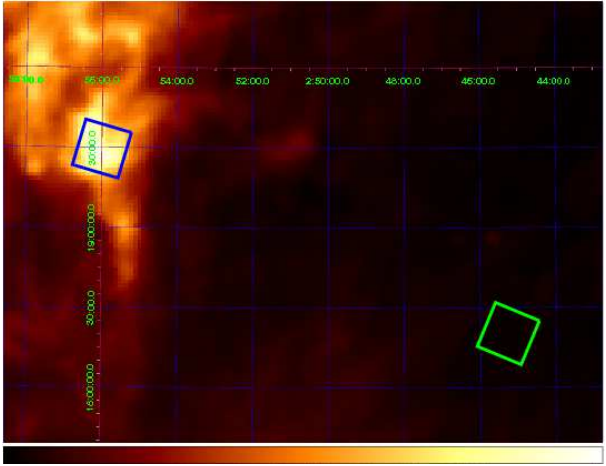}
\caption{\suzaku field of view for On-cloud (blue) and Off-cloud
  (green) pointings, overlaid on the IRAS 100$\mu$m survey image, which
  shows the bright emission from the MBM12 molecular cloud.
  \label{fig:FOV_iras100}}
\end{figure}

The molecular cloud MBM12 (Lynds 1457) was observed by \suzaku on
February 3-6, 2006 for a total of 231 ksec.  The nominal pointing
position was (RA, Dec) = $02^{h}56^{m}00^{s},
+19^{\circ}29^{'}24^{''}$\ (J2000) ($(l,b) = 159.2^{\circ},
-34.47^{\circ}$), corresponding to the most infrared-luminous and
thus densest portion of the molecular cloud.  Immediately
thereafter (on February 6-8, 2006) an ``off-cloud'' pointing was
obtained towards $02^{h}45^{m}16^{s}, +18^{\circ}20^{'}14.3^{''}$\
(J2000) ($(l,b) = 157.3^{\circ}, -36.8^{\circ})$, a position
$2.79^{\circ}$\ distant from the cloud, for 168 ksec.  We present data
primarily from the XIS instrument \citep{XIS06}, using the
back-illuminated CCD XIS1 which has the largest effective area at
energies below 1 keV.  The two fields of view are shown in
Figure~\ref{fig:FOV_iras100}, overplotted on the IRAS 100$\mu$m image.

We used version 0.7 of the \suzaku data processing pipeline for our
base dataset.  The cleaned v0.7 data are by default filtered to
exclude times within 436 seconds of \suzaku passing through the South
Atlantic Anomaly (SAA), and when \suzaku's line of sight is elevated
above the Earth's limb by less than $5^{\circ}$, or is less than
$20^{\circ}$\ from the bright-Earth terminator.  We decided to expand
this to exclude events with Earth-limb elevation angle less than
$10^{\circ}$, as there were some excess events in the 0.5-0.6 keV band
in the $5^{\circ}-10^{\circ}$\ range.  Finally, flaring pixels were
removed using the {\tt cleansis}\ tool with the default v0.7
parameters.  Although these are a small fraction of the total number
of pixels on the CCD, they contribute a sizable background.  In the
case of the on-cloud data for XIS1, just 1055 flickering pixels (out
of $\sim 10^6$\,total pixels) contributed $\sim 46\%$\ of the total
counts.

\begin{figure*}
\begin{center}
\FigureFile(80mm,80mm){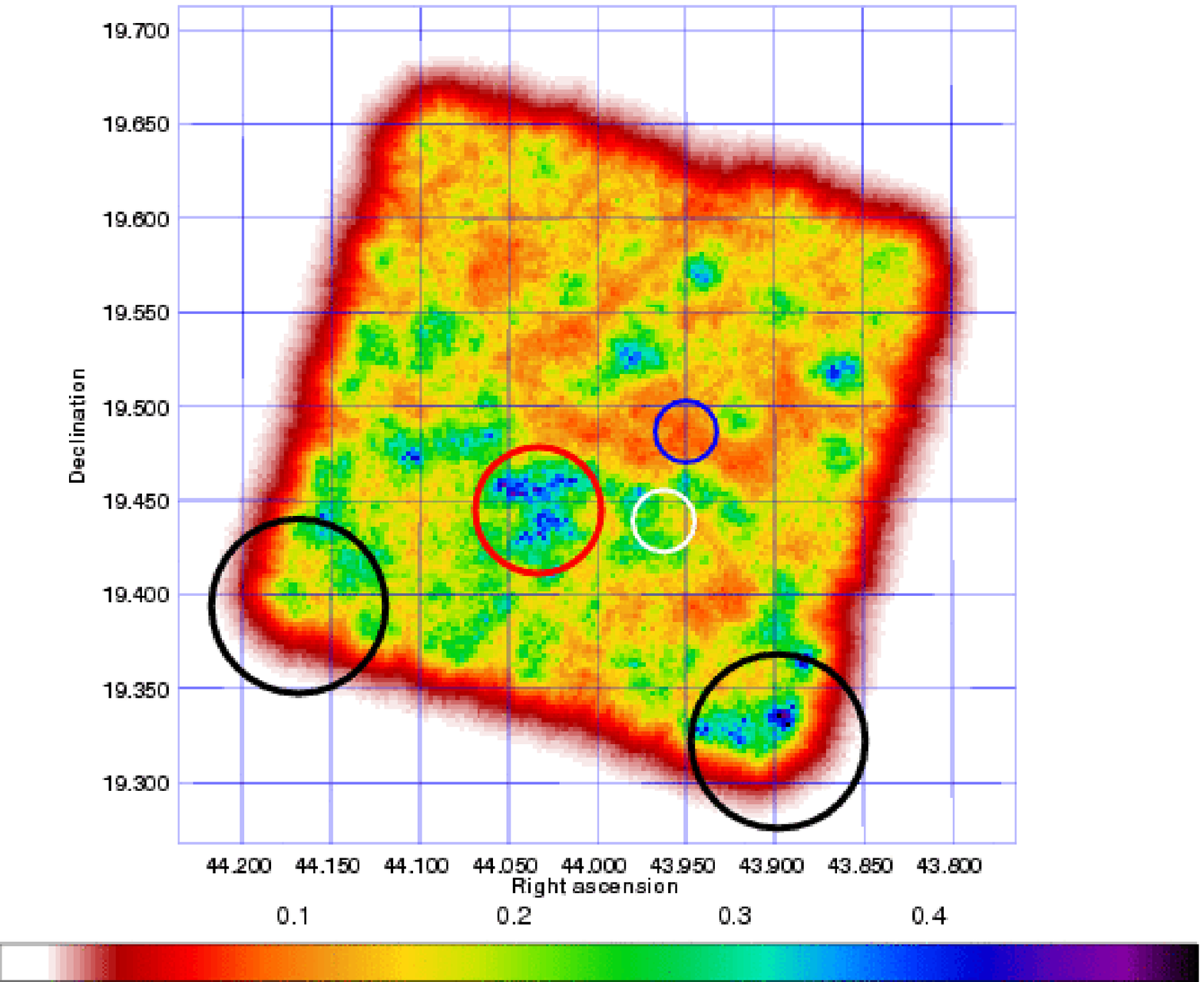}
\FigureFile(80mm,80mm){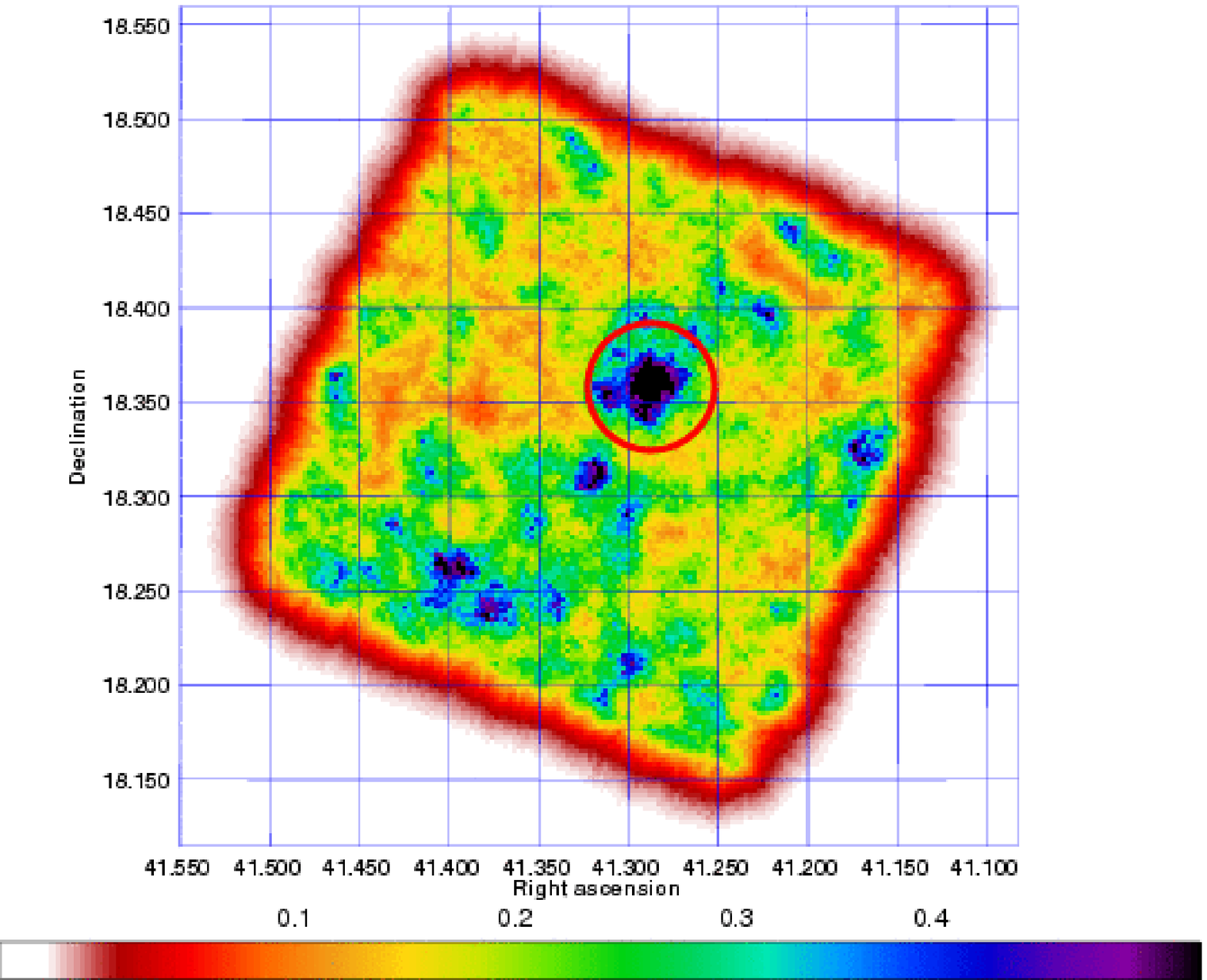}
\end{center}
\caption{[Left] MBM12 field of view between 0.4-1.0 keV, after
  smoothing.  Emission from XY Ari (red circle) can be seen, although
  the two weaker sources (blue, white circles) are not visible in this
  band.  The black circles indicate the position of the calibration
  sources. [Right] The off-cloud field of view, with an unidentified
  source (excluded in the analysis) marked with a $2'$\ radius red circle.
  \label{fig:image}}
\end{figure*}

The bright intermediate polar XY Ari was serendipitously included in
the on-cloud observations; this source will be discussed in a separate
paper.  XY Ari is sufficiently highly absorbed \citep{XYAri04} that no
photons below 1 keV are expected from the source.  However, a smoothed
image of the 0.4 - 1.0 keV band (see Figure~\ref{fig:image}[Left])
shows low-energy emission from XY Ari, likely due to the tail of the
CCD response curve.  We therefore excluded a $2'$\ radius region
around XY Ari (marked with a red circle in
Figure~\ref{fig:image}[Left]) in order to reduce this background.
This had the effect of substantially reducing the total background at
all energies while excluding only a small fraction of the total
$17.8'\times17.8'$\ field.  In addition, there were two other weaker
sources found which were also previously found in the \chandra
observation of MBM12.  These are marked in
Figure~\ref{fig:image}[Left] with a $1'$\ radius blue circle
(02:55:48, +19:29:12, J2000) and a white circle (02:55:51, +19:26:21,
J2000).  Both had hard spectra with no significant flux below 1 keV in
either the \suzaku or \chandra data.  In the off-cloud data
(Figure~\ref{fig:image}[Right]) we discovered a bright source at
02:45:09, +18:21:30 (J2000) (red circle) which does not appear in the
ROSAT All-Sky Survey or any other catalog.  We leave analysis of these
sources for a future paper, and concentrate on the diffuse soft X-ray
emission.


\subsection{Background}

As the goal of the observation was to extract the soft X-ray
background, which fills the field of view, other background components
cannot be estimated directly from the observation.  Therefore our
first focus was on understanding the importance of the three major
background components: particle contamination, scattered solar X-rays,
and solar wind charge exchange.  We did not exclude the corners of the
detectors which contained the onboard calibration sources in our
analysis, but instead fit these lines (which are all $>1$\,keV) as
part of our source and background models.

\subsubsection{Particle Background\label{ss:partbkgd}}

\suzaku is in a low-Earth orbit, so it is significantly shielded from
the particle background that strongly affects \xmm and \chandra.  The
effectiveness of this shielding is dependent upon the ``Cut-Off
Rigidity'' (COR) of the Earth's magnetic field, which varies as
\suzaku traverses its orbit.  During times with larger COR values,
fewer particles are able to penetrate to the satellite and to the XIS
detectors.  We considered using the default value (COR$>4$\,GV) but
finally chose to use a stronger constraint (COR$>8$\,GV) for both
observations, as the lowest background was desired.  This tighter
constraint eliminates 27\% (28.5 ksec) of the total on-cloud
observation time but 35\% of all the XIS1 counts.  After this cut, we
were left with 71.7 ksec of ``good'' time for the on-cloud observation
and 51.85 ksec for the off-cloud pointing.

Although it is reduced by the Earth's magnetic field, \suzaku still
has a noticeable particle background.  Fortunately, we can estimate
the background level quite accurately as a part of most observations
is spent observing the Earth at night, and these data are a good proxy
for the pure particle background.  Phenomena such as aurorae have been
observed to contribute X-rays to the Earth's night sky
\citep{Bhardwaj06}, but these processes tend to be transient and thus
are easily removed from the data.  We used $\sim 400$\,ksec of night
Earth observations from the SWG phase of the mission.  The data were
filtered to remove flares, and to ensure that \suzaku's line-of-sight
elevation from the Earth-limb was less than $-5^{\circ}$ while
observing the dark Earth.  We also required the same cut-off rigidity
constraint (COR$>8$\,GV) as used for the on- and off-cloud
observations when extracting the particle background spectrum.

\subsubsection{Scattered Solar X-rays\label{ss:ssx}}

As \suzaku orbits the Earth maintaining a fixed pointing, the column
density of atmosphere along the look direction varies rapidly.  Solar
X-rays can scatter off the atmosphere into the telescope, either via
Thompson scattering or by fluorescence \citep{SF93}.  Fluorescence of
oxygen atoms and molecules is our greatest concern, as it would give
rise to emission lines around 0.54 keV which could blend with the
O{\sc vii}\ line.  We modeled the Earth's atmosphere using the
NRLMSISE-00 empirical model \citep{Hedin90}, and combined this with
the \suzaku orbital parameters to calculate the total
solar-illuminated column density of oxygen atoms and O$_2$\ molecules
as a function of time.

\begin{figure}
\FigureFile(80mm,80mm){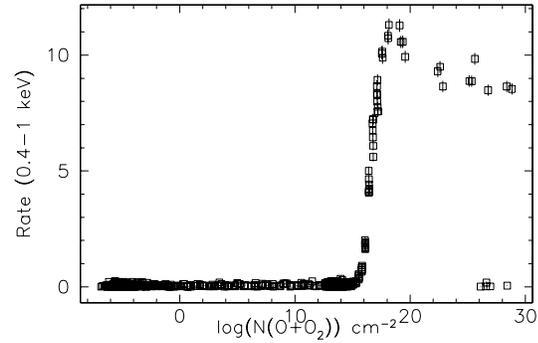}
\caption{The raw count rate in the 0.4-1.0 keV band for the on-cloud
  observation as a function of atmospheric column density of oxygen
  atoms and molecules.  The scattered solar X-ray contribution rises
  sharply at low elevations, where the atmospheric column density of
  oxygen is larger than $\sim 10^{15}$\,cm$^{-2}$,
  \label{fig:atmo_uncl}}
\end{figure}

\begin{figure}
\FigureFile(80mm,80mm){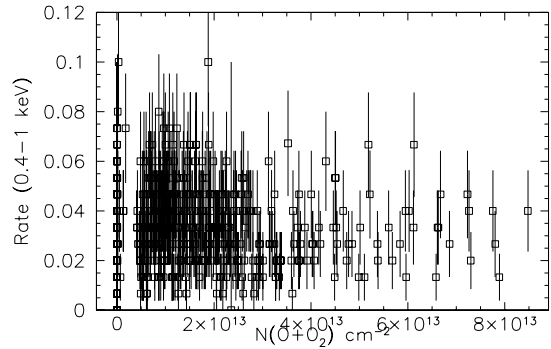}
\caption{The count rate after standard filtering in the 0.4-1.0 keV
  band as a function of atmospheric column density of oxygen atoms and
  molecules.  \label{fig:atmo_cl}}
\end{figure}

We then extracted the count rate as a function of time in the 0.4-1.0
keV band and compared this to the oxygen column density N$_{\rm O}$.
In Figure~\ref{fig:atmo_uncl} we show the correlation plot for the
uncleaned data.  When the illuminated atmospheric N$_{\rm O}$\ exceeds
$\sim 10^{15}$\,cm$^{-2}$, the count rate rises sharply due to
scattered solar X-rays.  We fit the data with N$_{\rm O}$\ between
$0-10^{17}$\,cm$^{-2}$\ to a linear model and found that they are well
described by the function $0.04 +
(1.18\pm0.01)\times10^{-16}$\,N$_{\rm O}$\,cts/s.  In
Figure~\ref{fig:atmo_cl} we show the same plot (with a linear
abscissa) after the standard filters are applied.  All the lines of
sight with column densities above $\sim 10^{14}$\,cm$^{-2}$\ have been
eliminated (at least for this observation) by the requirements that
the look direction be elevated by at least $10^{\circ}$\ and at least
$20^{\circ}$\ away from the bright Earth terminator.
Figure~\ref{fig:atmo_cl} shows that most times have either negligible
oxygen column or N$_{\rm O} \sim 10^{13}$\,cm$^{-2}$.  Applying our
linear fit, we see that the integrated contamination due to
fluorescent oxygen is less than $\sim 0.001$\,cts/s.  A similar result
holds for the off-cloud observation as well.  We therefore expect that
the scattered solar X-ray contribution to our data is negligible.

\subsubsection{Solar Wind Charge Exchange\label{ss:swcx}}

\begin{figure}
\FigureFile(70mm,70mm){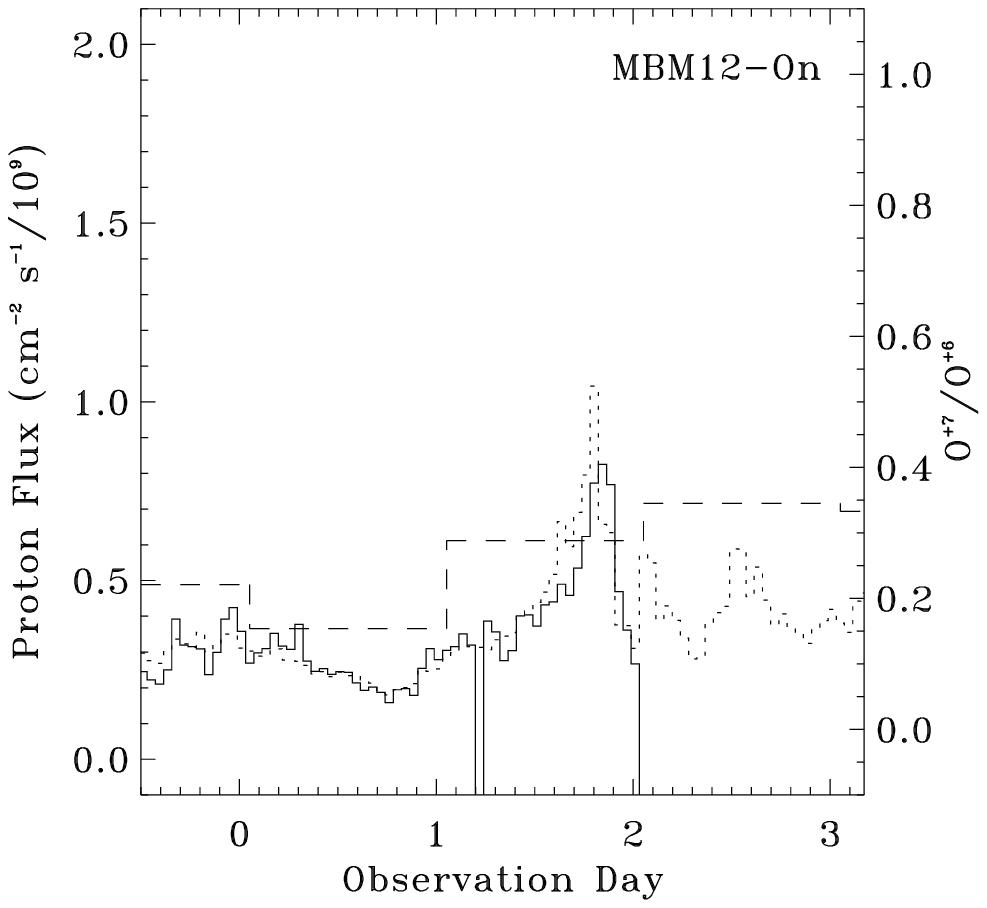}
\FigureFile(70mm,70mm){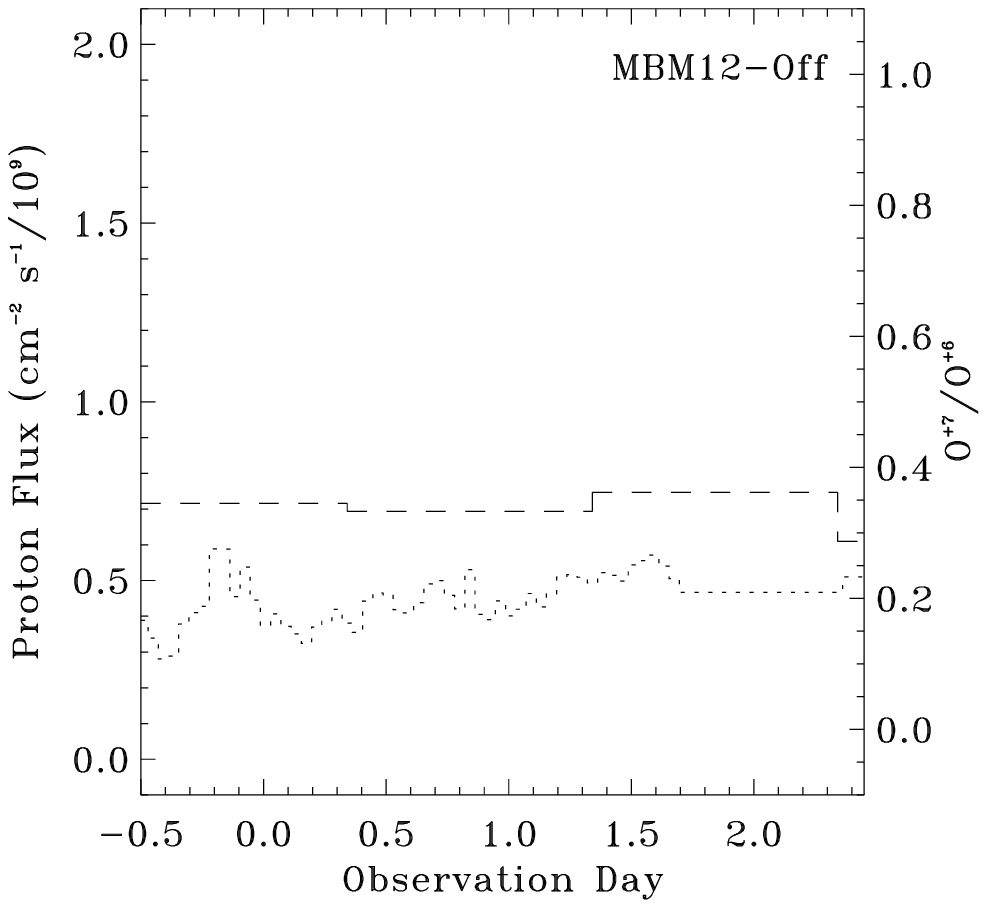}
\FigureFile(70mm,70mm){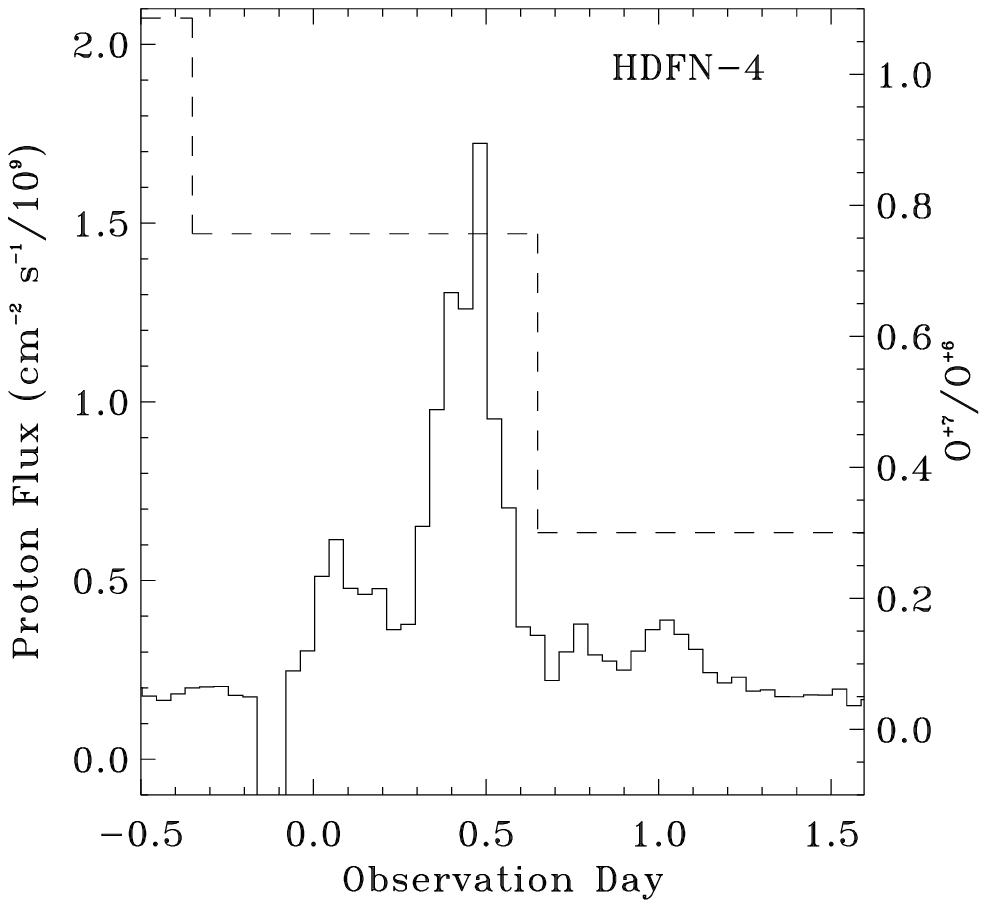}
\caption{[Top] The hourly average value of the solar wind proton flux
  in units of $10^9$\,cm$^{-2}$\,s$^{-1}$ from the ACE (solid line) and
  WIND (dotted line) satellites (the ACE and WIND data have not been
  corrected for the time of flight to \suzaku).  The daily average of
  the O$^{+7}$/O$^{+6}$\ ratio from ACE is shown by the dashed
  line. [Middle] Same, for the off-cloud observation. [Bottom] Same,
  shown for the Hubble Deep Field North \xmm observation.
  \label{fig:sw_params}}
\end{figure}

Diffuse soft X-ray emission can also be generated by ions in the solar
wind interacting with neutral interplanetary or geocoronal material.
However, the appearance and strength of emission lines emitted by SWCX
is poorly characterized.  It is expected that the density, velocity,
and ionization balance of the solar wind should correlate with the
variable SWCX contribution, which is largely from the geocorona, but
there may also be a more nearly constant component of the SWCX
emission from interplanetary space.  Figure~\ref{fig:sw_params} shows
some of the relevant values for the two \suzaku observations and, for
comparison, during an instance of strong SWCX emission seen by \xmm.

\citet{SCK04} analyzed an \xmm observation of the Hubble Deep Field
North that showed substantial SWCX emission.  That observation
occurred during a period characterized by a strong solar proton flux
(in only a few percent of observations are stronger fluxes seen), as
well as high O$^{+7}$/O$^{+6}$, but low O$^{+8}$/O$^{+7}$.  The proton
speed was low, $\sim350$ km\,s$^{-1}$.  Enhanced O{\sc vii}
($7.39\pm0.79$\,LU) and (despite the low O$^{+8}$/O$^{+7}$\ ratio)
O{\sc viii} ($6.54\pm0.34$\,LU) emission was seen during this
observation, along with a number of other species.  In contrast, our
\suzaku observations were done at a time characterized by a moderate
proton flux in the solar wind, with the exception of one short period
of the on-cloud observation.  The O$^{+7}$/O$^{+6}$ ratio during both
observations was close to the mean ratio for the solar wind.  The
proton speed during both \suzaku observations was exceptionally low,
typically $<350$ km\,s$^{-1}$, which is seen in only a few percent of
observations.

Nonetheless, there were some similarities in the solar wind parameters
during the \xmm observation that showed strong SWCX contamination and
our \suzaku observations.  In both, the solar wind was slow and dense.
However, the peak proton flux during the \suzaku observations was less
than half the proton flux responsible for the SWCX emission in the
\xmm observation, and the mean proton flux is even lower.  Further,
the \xmm line of sight observed through the densest portion of the
Earth's magnetosheath, whereas the \suzaku observations are through
the flanks of the magnetosheath, thus further reducing the target
neutrals with which the solar wind produces the X-ray emission.
Unfortunately, our understanding of how the solar wind
characteristics, satellite orbits, and observing directions interact
to generate observed SWCX emission is still quite limited.
Nonetheless, the combination of the solar wind strength and the look
direction, as well as the fact that the ion ratios are close to the
mean values, suggests that whatever the SWCX surface brightness was
during our \suzaku observations, it is typical for observations of the
diffuse soft X-ray background.

\subsection{XIS Response}

For this work, we focused on the back-illuminated XIS1 detector, which
has the largest effective area of low energy X-rays.  As these
observations were performed early in the mission, very little
degradation of the CCD response had occurred.  Unfortunately, however,
the time- and space-varying contamination layer which was discovered
early in the mission has complicated observations at low energies
\citep{XIS06}.

We calculated the XIS detector effective area using the tool {\tt
xissimarfgen}, which includes both the time and spatial effects of the
contamination layer.  We assumed a field-filling source, and used a
detector mask which removed the bad pixel regions, along with the
region excluded due to the bright source XY Ari.  The model of the
contamination layer is based on in-flight observations and has its own
uncertainties.  Combining these with other known sources of systematic
error, we estimate that at 0.6 keV there is a $\sim 13$\% systematic
error on the final effective area$\times$solid angle product, in
addition to the given statistical errors.

\section{Results}

\subsection{On-Cloud Emission}

Although MBM12 absorbs almost all distant emission below 0.7 keV, we
cannot say if the foreground low energy emission is local to the solar
system or tens of parsecs away.  Our first goal, however, is to simply
model the spectrum seen towards MBM12 since this is likely the
'darkest' high latitude line of sight in the Galaxy at soft X-ray
energies.

\subsubsection{Raw Count Model\label{ss:simple}}

\begin{figure}
\FigureFile(80mm,80mm){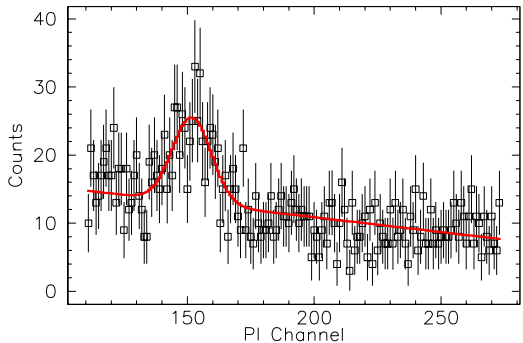}
\caption{The on-cloud spectrum between 0.4-1.0 keV (channels 110-273)
  in channel units.  The best fit line plus Gaussian is shown.
 \label{fig:mbm12_simplespec}}
\end{figure}

The data clearly showed a feature near 0.56 keV, so we began by simply
fitting a linear continuum plus a Gaussian to the observed count rates
(with no background subtraction) for the on-cloud data on XIS1 between
0.4-1.0 keV in PI channels.  This approach is admittedly simplistic,
but it gives a baseline measurement, useful when comparing to a more
complicated physical model.  Figure~\ref{fig:mbm12_simplespec} shows
the best-fit result, which has $229^{+34}_{-32}$\,counts in the line
and a centroid at PI channel $151.6\pm1.0$\, or $553.3\pm3.9$\,eV
(using 3.65 eV/channel).  Taking into account the area removed for bad
pixels and the effects of the optical blocking filter contamination
with a thickness appropriate for Day 209, \suzaku's effective area
$\times $\ solid angle product at 0.553 keV is 16,875
cm$^2$\,arcmin$^2$.  With a total ``good time'' of 71.7 ksec, we get a
total surface brightness of $2.23\pm0.32$\,LU.  We also put a
2$\sigma$\ upper limit on any O{\sc viii}\ line (at 0.653 keV) of 20
counts (0.11 LU) using this method.  Since the total number of counts
is rather small, in this and all subsequent fits, we used the
maximum likelihood Cash statistic \citep{Cash79}.

\subsubsection{Physical Model}

To expand upon these simple results, we then considered a more
realistic physical model which explicitly included the detector
background along with known astrophysical sources.  We restricted the
energy range to 0.4-7 keV, as above 7 keV the particle background
rises sharply.  The background was fit to the night Earth data (see
\S\ref{ss:partbkgd}) using a model consisting of the sum of a
power-law, a constant, and the five emission lines expected in this
energy range (see Table 6.2 in the Suzaku Technical
Description\footnote{http://heasarc.gsfc.nasa.gov/docs/suzaku/prop\_tools/suzaku\_td/}).
The emission lines were modeled as Gaussians (see
Table~\ref{tab:bglines}).
\begin{table*}
\caption{Instrumental Emission Lines\label{tab:bglines}}
\begin{center}
\begin{tabular}{llllllll}
\hline \hline 
Ion & & \multicolumn{3}{c}{Night Earth} & \multicolumn{3}{c}{On-Cloud} \\
    &E$_{\rm lab}$& E   & FWHM & S.B &   E  & FWHM & S.B.  \\
    & keV         & keV & keV  & LU  & keV  & keV  & LU   \\ \hline
Al K& 1.49        &1.48 & 0.044&0.28 &1.48 & 0.044&0.28  \\
Si K& 1.74        &1.74 & 0.053&0.36 &1.74 & 0.053&0.36  \\
Au M&2.123       &2.13 & 0.170&0.34 &2.13 & 0.170&0.34  \\
Mn K$\alpha$&5.90&5.88 & 0.079&15.16&5.88 & 0.079&15.11 \\
Mn K$\beta$&6.49  &6.47 & 0.096&5.35 &6.47 & 0.096&5.38  \\ \hline
\end{tabular}
\end{center}
\end{table*}
Note that the best-fit energies agree with the laboratory energies to
within 1\%.  The variation in the FWHM is not completely understood,
but the large value at 2.13 keV is probably due to the multiple lines
found in the Au M complex.  The power-law term (with best-fit $\Gamma
= 1.02$\ and amplitude 0.011 counts s$^{-1}$keV$^{-1}$\ at 1 keV) and
the constant (amplitude 0.00723 counts s$^{-1}$keV$^{-1}$) were not
folded through the effective area curve.  These two terms account for
the observed particle background, after the COR$>8$\,GV and
ELV$>10^{\circ}$\ filters.

The source model included two absorbed broken power-laws to account
for the cosmic X-ray background (CXRB) and an absorbed bremsstrahlung
plus Fe line for the remaining XY Ari emission (see below).  A
Gaussian line was added to represent the blended N{\sc vi} triplet and
C {\sc vi} Ly$\beta$\ line, and a final Gaussian was included to
represent the O{\sc vii}\ emission.  The broken power-law components
fit the composite total AGN spectrum, giving a slope of 1.4 above 1.2
keV and steepening significantly below 1 keV as observed by ROSAT and
\chandra \citep{Hasinger93,Mushotzky00}.  The first broken power-law
was fixed with $\Gamma_1 = 1.54$, $\Gamma_2 = 1.4$, $E_b = 1.2$, and a
normalization of $5.70$\,ph\,cm$^{-2}$\,sr$^{-1}$\,s$^{ -1}$\ at 1
keV.  The second broken power-law used fixed values of $\Gamma_1 =
1.96$, $\Gamma_2 = 1.4$, $E_b = 1.2$, but the normalization was
allowed to vary below its nominal value of
4.90\,ph\,cm$^{-2}$\,sr$^{-1}$\,s$^{ -1}$\ at 1 keV; our best-fit
value was $2.53\pm0.36$\,ph cm$^{-2}$\,sr$^{-1}$\,s$^{-1}$\ at 1 keV.
Both components were assumed to be absorbed with column density
$N_{\rm H} = 4\times10^{21}$\,cm$^{-2}$, using the value for MBM12
found in \citet{Smith05}.

Despite the exclusion of the region within $2'$\ of XY Ari from the spectrum,
the source is so bright that its scattered emission contributes
significantly to the overall spectrum above 1 keV.  This contribution
was modeled as absorbed bremsstrahlung emission with an additional
iron line.  The best-fit value had N$_{\rm H} =
5.5\times10^{22}$\,cm$^{-2}$, $kT = 200$\,keV, and a total absorbed
surface brightness (0.4-7 keV) of $1.25\times10^{-7}$\,erg
cm$^{-2}$\,s$^{-1}$\,sr$^{-1}$.  The best-fit Fe line was at 6.98 keV,
with FWHM 0.4 keV and surface brightness 1.3 LU.  We note that, while
this model fits the X-ray spectrum of XY Ari reasonably well, we do
not claim it is a correct physical model of the emission.  An initial
fit to the XY Ari data itself (using data from the central $2'$)
showed that the scattered flux is $\sim 22$\% of the total source
flux, in agreement with the expected value (19\%) based on the XRT PSF
after excluding the central $2'$.  The best-fit temperature was lower
($kT = 45_{-11}^{+19}$\,keV), and the absorption higher (N$_{\rm H} =
6.7\times10^{22}$\,cm$^{-2}$).  The column density found for XY Ari is
more than an order of magnitude larger than the MBM12 value, although
it is similar to the value found by \citet{Littlefair01}, who used
K-band spectroscopy to determine that XY Ari's secondary is an M0V
star, with an $A_V = 11.5\pm0.3$, corresponding to a hydrogen column
density of $2.2\times10^{22}$\,cm$^{-2}$.  However, \citet{Luhman01}
showed that most stars within the MBM12 cloud have $A_V < 2$, while
background stars generally have values between $A_V = 3-8$. The origin
of the discrepancy between these values and \citet{Littlefair01} is
unknown, but may be due to a inadequate model for the X-ray spectrum
of XY Ari.  The absorbing material may be near XY Ari itself, although
it is also possible that MBM12 has a larger column density along this
line of sight that the average value we assumed.  This will not affect
our results since our model already has little to no flux in the
0.5-0.7 keV band from beyond MBM12.  In any event, a more detailed
analysis of the XY Ari data is in progress, and we are certain the
effect on the continuum below 1 keV is small.

The contribution from the Local Hot Bubble itself is normally modeled
as a thermal plasma in CIE with T$\sim 10^6$\,K.  However, most of the
LHB emission is in the 0.1-0.3 keV bandpass, where \suzaku has some
effective area but is not yet accurately calibrated.  With our lower
energy limit of 0.4 keV, the only strong lines expected from the LHB
are from the N{\sc vi}\ triplet at $\sim 0.43$\,keV, along with C{\sc
vi}\,Ly$\beta$\ emission at the same energy.  We therefore included a
single Gaussian to represent these lines, and ignored the continuum since
this is negligible in a thermal plasma with T$\sim 10^6$\,K.  The
best-fit position was $0.42\pm0.03$\,keV, with FWHM 0.058 keV and
surface brightness $2.4^{+2.2}_{-0.60}$\,LU.

\begin{figure}
\FigureFile(80mm,80mm){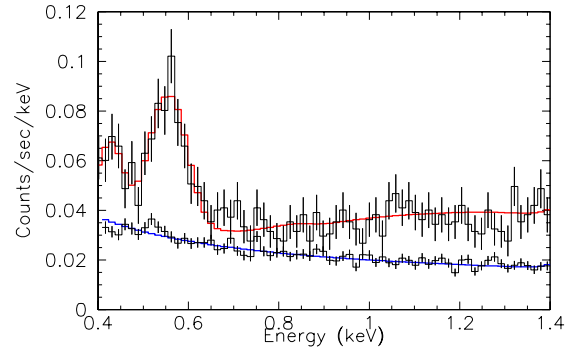}
\caption{The on-cloud spectrum with the best-fit model (red line) with
  the background spectrum and best-fit model (blue line) between 0.4-1.4 keV
 \label{fig:oncloudspec}}
\end{figure}

The final term was a Gaussian to represent the oxygen emission.  The
best-fit parameters put the line at $0.556\pm0.003$\,keV, with FWHM
0.071 keV and a total surface brightness of $3.53\pm 0.26$\,LU.  The
line position is nearly identical to that found in \S\ref{ss:simple},
while the surface brightness is increased by 60\%.  In this model, the
continuum (due to particle background, the tail of the CCD response,
and the absorbed CXRB) is very low at the O{\sc vii} line, as opposed
to the simple model which assumed a flat continuum under the line.
The best-fit spectrum, including the background night Earth data, is
shown in Figure~\ref{fig:oncloudspec}

In the simple model, we were able to put a 2$\sigma$\ upper limit on
any O{\sc viii} contribution by adding a delta function at the
expected position of an O{\sc viii} line.  Likewise here we added a
delta function to the model at 0.653 keV, to represent the O{\sc viii}
Ly$\alpha$\ line. The best-fit result is a marginal detection of a
feature with surface brightness $0.24\pm0.1$\,LU, which when included
in the model reduces the O{\sc vii}\ surface brightness to
$3.34\pm0.26$\,LU.  The O{\sc viii}/O{\sc vii} surface brightness
ratio is then $7.2\pm3.0$\%.  \citet{Smith05} noted that the O{\sc
vii} $n=3\rightarrow1$\ transition (at 0.666 keV) line can contribute
as much as 6\% of the flux of the main O{\sc vii} $n=2\rightarrow1$\
triplet.  Although we do not claim this as a detection, it seems more
likely that this emission is from this O{\sc vii} line and not O{\sc
viii}.

\subsection{Off-Cloud Emission}

The off-cloud observations were taken immediately following the
on-cloud data and as shown in \S\ref{ss:swcx}, the solar wind
parameters were relatively stable during this period. So assuming the
SWCX contribution is stable, we can use the difference between these
observations as an estimate of distant Galactic disk and halo
emission.

\begin{figure}
\FigureFile(80mm,80mm){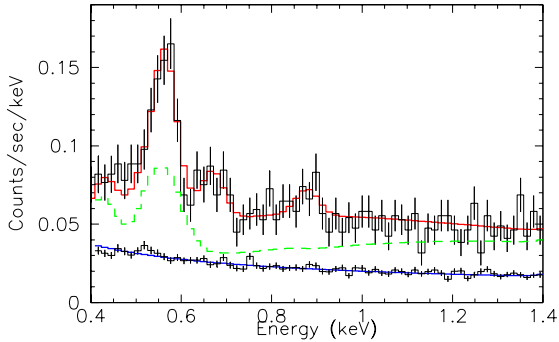}
\caption{The off-cloud spectrum (top) and best-fit model (red line),
  showing the three features at O{\sc vii}, O{\sc viii}, and an
  unknown feature at 0.876 keV.  The green dashed curve shows the
  best-fit model for the on-cloud spectrum (data points omitted for
  clarity). The best-fit model for the night Earth data (bottom curve)
  is in blue.
  \label{fig:mbm12_distant}}
\end{figure}

We assumed the ``background'' (actually a foreground in this case) for
the off-cloud spectrum is the same as the on-cloud spectrum without
the contribution from XY Ari.  We assume that the ``distant'' emission
originates beyond most of the Galactic gas (with N$_{\rm H} =
8.7\times10^{20}$\,cm$^{-2}$) seen in this direction.
Figure~\ref{fig:mbm12_distant} shows the best-fit to the off-cloud
spectrum between 0.4-1.5 keV.  We added two Gaussian lines to the
model to represent the ``distant'' emission from O{\sc vii} and O{\sc
viii}, as well as a third (with FWHM set to 0 to force the fit to
reflect a single line, rather than a very wide blend) to fit the
excess between 0.85-0.9 keV.  The ``local'' Oxygen emission lines
measured in the on-cloud observation were also included in this fit,
so these new lines measure only the ``distant'' component.  The fit
parameters are given in Table~\ref{tab:mbm12_distant}.  The O{\sc vii}
$n=3\rightarrow1$\ line (at 0.666\,keV) may contribute up to 6\% of
the O{\sc vii} emission ($\lesssim 0.14$\,LU) to the O{\sc viii}\
feature.  This line could also be responsible in part for shifting the
best-fit line energy above 0.653\,keV, the energy of the O{\sc viii}
Ly$\alpha$\ line.  Nonetheless, the majority of the emission at
0.668(6)\,keV must be from O{\sc viii}, as it is the only strong line
near this energy.

The O{\sc viii} detection indicates that the distant plasma is either
hotter or more out of equilibrium than the LHB plasma.  In either
case, the 0.7-1.3 keV range may contain relatively weak emission from
Neon and Fe L shell lines.  To put a limit on any such emission, we
added a delta function to the model at 0.826 keV (15.01\AA), where the
strongest Fe feature, from Fe {\sc xvii}, would be expected.  The
2$\sigma$\ upper limit on the Fe XVII line is 0.19 LU.  This is not
unexpected, since the expected surface brightness in this line is only
$\sim 3\%$\ of the O{\sc viii} value assuming an Fe/H abundance of
$3.24\times10^{-5}$\ and a temperature of $\sim 2\times10^6$\,K (see
\S\ref{ss:offcloud}).  This reinforces the confusing nature of the
unknown feature at 0.876 keV, since its origin is therefore almost
certainly not Fe L shell emission.

\begin{table}
\caption{Best-fit parameters for distant
  emission, with 1$\sigma$\ errors\label{tab:mbm12_distant}}
\begin{tabular}{llll}
\hline \hline
Ion         & Energy     & FWHM  & Flux \\
            & keV        & keV   & LU  \\ \hline
O{\sc vii}  & $0.562(4)^a$  & $<0.001$& $2.34\pm0.33$ \\
O{\sc viii} & $0.668(6)$  & 0.02(2)   & $0.77\pm0.16$ \\
Unknown     & $0.876(9)$  & 0         & $0.26\pm0.08$ \\ \hline
\end{tabular}

${^a}$\ Value in parentheses shows error on the last digit
\end{table}

\section{Discussion \& Conclusions}

\subsection{On-Cloud}

By observing both MBM12 and a nearby ``off-cloud'' field in quick
succession with a low-background imaging X-ray spectrometer, we can
reliably measure both the foreground and distant emission.  Our
on-cloud result indicates a rather high value for the ``local''
diffuse O{\sc vii} surface brightness of about 3.5 LU.  This emission
is almost certainly generated in front of MBM12 since the cloud is
optically thick, with transmission $<3\%$ at O{\sc vii}.  Assuming the
O{\sc vii} surface brightness behind MBM12 is the same as the value
given in Table~\ref{tab:mbm12_distant}, the background contribution to
the on-cloud emission is $<0.06$\,LU.  As first suggested by
\citet{Cox98}, the most likely sources for the foreground emission are
either the LHB, SWCX, or both.

Although larger than expected for standard LHB models, the local O{\sc
vii} surface brightness is still lower than most measured values
towards high-latitude sightlines.  \citet{XQC02} found $4.8\pm0.8$\,LU
of O{\sc vii} towards a 1 sr region at high latitude, while the
2$\sigma$\ upper limit set toward MBM12 by \citet{SnowdenMBM12} with
ROSAT is equivalent to 7.1 LU at O{\sc vii}.  ASCA observations
towards high-latitude sightlines found a surface brightness of
$2.3\pm0.3$\,LU in O{\sc vii}, although this was model dependent
\citep{Gendreau95}.

\citet{Smith05} did find a lower value for the local O{\sc vii}
surface brightness ($1.92^{+0.61}_{-0.60}$\,LU, statistical errors
only) when observing MBM12 with \chandra.  However, \chandra's
particle background is both larger and more uncertain than \suzaku's.
In addition, a large solar flare occurred during the \chandra
observation, and \citet{Smith05} noted that this increased the
uncertainty.  \citet{Smith05} also measured an O{\sc viii} surface
brightness of $2.35^{+0.59}_{-0.43}$\,LU, which they attributed
largely to SWCX particle background (along with an unknown portion of
the O{\sc vii} emission).  We find a marginal detection of
$0.24\pm0.1$\,LU of O{\sc viii} towards MBM12.  It seems likely that
including the poorly-understood systematic errors on the \chandra
O{\sc vii} result would bring it into agreement with the \suzaku
measurement, although we cannot say with precision how much of either
is due to SWCX and not the LHB.  In contrast, the O{\sc viii} seen
with \chandra is almost certainly due to SWCX \citep{Smith05}.

It is difficult to generate the observed O{\sc vii} surface brightness
from a solar abundance LHB model consistent with ROSAT and other
observations, as a model that generates the observed O{\sc vii}
emission will predict too much 1/4 keV band emission.  As described in
\citet{Smith05}, the dominant source of O{\sc vii} flux in a thermal
plasma is collisional excitation of O$^{+6}$\,ions, unless the plasma
is substantially out of equilibrium.  We are, however, reasonably
certain the LHB is near equilibrium.  The strongly recombining model
proposed by \citet{BS94} was ruled out by \citet{Smith05}.  They
showed that the 2$\sigma$\ upper limits to the O{\sc vii} and O{\sc
viii} surface brightnesses from the \chandra MBM12 observation,
combined with upper limits on the O{\sc vi} surface brightness from
FUSE \citep{Shelton03} and the ROSAT All-sky surface 1/4 keV band
brightness \citep{Snowden95} excluded this type of model.  Conversely,
\citet{EC93} considered models of the LHB arising from a strongly
ionizing plasma, created by a recent supernova.  They concluded that
generating the observed soft X-ray emission required an
unrealistically high ambient ISM pressure (p/k
$>3\times10^4$\,cm$^{-3}$K).  In addition, these types of models
typically predict significant amounts of high-velocity O{\sc vi},
which have not been observed \citep{Shelton02}.

We therefore consider models in collisional equalibrium.  A CIE plasma
at $10^6$\,K, the ``typical'' temperature for the LHB (SMV93), has an
O{\sc vii} triplet emissivity of $\Lambda = 5.7\times10^{-16}$\,ph
cm$^3$\,s$^{-1}$, assuming the oxygen abundance relative to hydrogen
is $8.51\times10^{-4}$ \citep{SmithAPEC}.  This rises rapidly with
temperature, reaching $1.5\times10^{-15}$\,ph cm$^3$\,s$^{-1}$ at
$1.2\times10^6$\,K, and peaking (for T $=2\times10^6$\,K) at
$6.4\times10^{-15}$\,ph cm$^3$\,s$^{-1}$.  However, the $2\sigma$\
upper limit on the O{\sc viii}/O{\sc vii} ratio of 13\% puts an upper
limit on the equilibrium temperature of $1.7\times10^6$\,K (where
$\Lambda = 4.8\times10^{-15}$\,ph cm$^3$\,s$^{-1}$).

Using these values, and assuming a constant density and temperature
throughout the LHB, we can express the total surface brightness of
O{\sc vii} as:
\begin{equation}
L_{S}(\hbox{O{\sc vii}}) = {{1}\over{4\pi}} R_{\hbox{\small LHB}} 
  {{n_e^2}\over{1.2}} \Lambda_{\hbox{\small {O{\sc vii}}}}
\label{eq:sbsimp}
\end{equation}
where L$_S$\ is in LU, $n_e$\ is the electron density and
R$_{\hbox{\small LHB}}$\ is the bubble radius.  This assumes that
hydrogen and helium are fully ionized, so $n_e \approx 1.2 n_H$.
Taking our lower value of 2.3 LU of O{\sc vii},
Equation~(\ref{eq:sbsimp}) gives $n_e^2 R_{\hbox{\small LHB}} =
0.020$\,cm$^{-6}$pc at $10^6$\,K, or 0.0075\,cm$^{-6}$pc at
$1.2\times10^6$\,K.  We can obtain a lower limit of
0.0023\,cm$^{-6}$pc on this value using our upper limit of
$T=1.7\times10^6$\,K.  This final value is similar to the emission
measure found by SMV93 (0.0024\,cm$^{-6}$pc), although at a
significantly higher temperature.  Assuming R$_{\hbox{\small LHB}} =
100$\,pc, we require electron densities of 0.014, 0.0087, or 0.0048
cm$^{-3}$, and a pressure of $p/k =3.0, 2.2,$\ or
$1.7\times10^4$\,cm$^{-3}$K at $T = 10^6$\,K, $1.2\times10^6$\,K, or
$1.7\times10^6$\,K, respectively.

Interestingly, \citet{Cox05} found that the midplane pressure required
to support the various layers of the Galaxy ({\it e.g.}\ cold and warm
H{\sc i}, diffuse H{\sc ii}, etc) is $2.2\times10^4$\,cm$^{-3}$K, in
agreement with our value at $T = 1.2\times10^6$\,K.  In addition,
\citet{Snowden00} used X-ray shadows (such as those created by MBM12)
seen in the ROSAT All-sky Survey observations at 1/4 keV to measure
the temperature of the local diffuse soft X-ray component.  They also
found a best-fit temperature of $1.2\times10^6$\,K, although this is
based on the \citet{RS77}(RS77) plasma code.  In particular, using
this temperature and the pressure with the 1993 update to the RS77
plasma code (using solar abundances) would predict $\sim 3\times$\ the
observed 1/4 keV band surface brightness seen by ROSAT.  This could
perhaps be explained if the Si, Fe, and other high-Z elements that
create the 1/4 keV band emission were depleted relative to oxygen in
the LHB; more modelling is needed to test this hypothesis.  

Despite the suggestive agreement in temperature and pressure described
above, there are issues in other wavebands.  \citet{Hurwitz05} has
placed a 95\% upper limit on the emission measure of a local $10^6$\,K
component of 0.0004 cm$^{-6}$pc, based on CHIPS observations of
diffuse EUV iron lines and assuming a solar abundance for iron.  Even
if iron is fully depleted, they still find a 95\% upper limit of $\sim
0.005$\,cm$^{-6}$pc for any CIE model with T$<1.6\times10^6$\,K, based
on the non-detection of O{\sc v}\ and O{\sc vi}\ lines near
171-173\AA.

The fully-depleted \citet{Hurwitz05} upper limit disagrees with our
value of 0.0075 cm$^{-6}$\,pc by a factor of at least 50\%.  In
addition, the solar-abundance CHIPS limit strongly disagrees with the
value found by SMV93 and from the ROSAT All-Sky Survey in the 1/4 keV
band ($\sim 0.0018 - 0.0058$\,cm$^{-6}$pc) which also assumes solar
abundances \citep{Snowden98}.  \citet{Hurwitz05} noted these
discrepancies and suggested that some depleted abundance pattern might
exist that brings the X-ray and EUV observations into agreement.
Nonetheless, as it stands the fully-depleted CHIPS limits suggest that
at least a third of the O{\sc vii} we detect is not from the LHB.  One
possibility is that this emission comes from SWCX.  More analysis of
the solar wind data will be needed to determine if the observations
were truly done during a period of relative quiescence; for example,
the absolute O$^{+7}$\ and O$^{+8}$\ fluxes can be derived from ACE
data with additional effort.  While Figure~\ref{fig:sw_params}, based
on the automatically processed ACE data, does not show any signs of
increased oxygen flux, more data are needed to confirm this.

\subsection{``Distant'' emission\label{ss:offcloud}}

Figure~\ref{fig:sw_params} shows that the solar wind conditions were
similar during both observations, and the LHB intensity is not
expected to change over an angle of less than $3^\circ$.  If both
oxygen lines are from an unabsorbed plasma in CIE, the O{\sc
viii}/O{\sc vii} ratio ($0.33\pm0.08$) implies $T =
(2.2^{+0.1}_{-0.2})\times10^6$\,K.  At this temperature, the predicted
emission measure is $(1.9\pm0.3)\times10^{-3}$\,cm$^{-6}$pc using
ATOMDB v1.3.1 emissivities \citep{SmithAPEC}.  If, as is more likely,
the plasma is behind the Galactic hydrogen layer ($N_{\rm H} =
8.7\times10^{20}$\,cm$^{-2}$), then the unabsorbed O{\sc viii}/O{\sc
vii} ratio would be $0.26\pm0.06$.  In this case, $T =
2.1\pm0.1\times10^6$\,K and the emission measure is
$(4.0\pm0.6)\times10^{-3}$\,cm$^{-6}$pc.  In either case, our results
consistent with previous measurements of distant hot halo gas.
However, our result does not touch on the question of whether the halo
has one \citep{Pietz98} or two \citep{KS00} dominant temperatures;
further \suzaku observations will be necessary to address this
question.

The line at 0.876 keV is a mystery, although we stress it is at best a
$ 3\sigma$\ detection.  Between 0.7-1.3 keV, the strongest emission
lines in a collisional plasma are typically from neon or iron.  The
closest strong neon line to 0.876 keV is the Ne{\sc ix} forbidden
line, but this would require a 5\% gain error in the XIS1.  The oxygen
lines at lower energies show $<2\%$\ gain shift, and the calibration
lines at higher energies (see Table~\ref{tab:bglines}) have less than
1\% gain shift.  The strongest iron lines near this energy are from
$2p^43d\rightarrow2p^5$\ transitions in Fe {\sc xviii}, but any
identification with an Fe line is problematic since many other lines
of Fe{\sc xviii}\ (such as the $2p^43s\rightarrow2p^5$\ line at 0.775
keV) would also be expected.  In particular, the 2$\sigma$\ upper
limit on the Fe{\sc xvii} 0.826 keV line of 0.19 LU strongly limits
any Fe line identification for the line at 0.876 keV.  It is possible
it is an as-yet unidentified weak instrumental line, although this
raises the question of why it is not present in the on-cloud data.

Intriguingly, the Lyman limit for O{\sc viii}\ is 0.8704 keV, so it is
possible that this is not a line, but rather a recombination edge
resulting from cool electrons interacting with O$^{+8}$\ ions.  If so,
we wonder at the origin of the O$^{+8}$\ ions--are they local to the
Solar system due to a sudden change in the solar wind during the
off-cloud observation, or from a distant recombining plasma?  As more
data from ACE and \suzaku becomes available, we may be able to answer
this question.

\section{Acknowledgments}

We would like to thank the \suzaku operations team for their support
in planning and executing these observations, along with Keith Arnaud
and John Raymond for helpful conversations.  JPH acknowledges support
from NASA grant NNG05GP87G.


\end{document}